\documentclass[
   ,final
  ]
  {aipproc}

\layoutstyle{6x9}

\begin{document}
\title 
      {Elliptic Flow from Au+Au Collisions at $\sqrt{s_{NN}} = 200$ GeV}

\author{A. Tang for the STAR collaboration}{
   address={Brookhaven National Laboratory, USA \& The National Institute for Nuclear Physics and High Energy Physics, Netherlands}
} 

\copyrightyear  {2003}

\begin{abstract}

This paper presents results of elliptic flow measurements at moderate high 
transverse momentum in Au+Au collisions using the STAR detector at RHIC. Sizable $v_{2}$ 
is found up to $7$ GeV/c in transverse momentum. Non-flow effects are discussed 
comparing  correlations in p+p collisions and  Au+Au collisions. $v_{2}$ from two-, four- and
six-particle cumulant are shown and discussed.

\end{abstract}

\maketitle

\vspace{-1.2 cm}

\section{Introduction}

The azimuthal anisotropy of produced particles at large transverse
momentum in non-central heavy ion collisions is one of several promising
observables for probing the early partonic phase~\cite{Olli92,Sorge}, 
and in particular, this phenomenon can provide insights into partonic 
energy loss in the relevant medium~\cite{MGyulassy2001}. 
Partonic energy loss increases systematically with 
increasing initial medium density and thus provides an important 
constraint on the initial parton densities. At moderate to large 
transverse momentum, the jet fragmentation process introduces a 
genuine correlation among fragmentation products, which is expected to 
be the dominant non-flow source~\cite{Olli00} that can complicate 
the interpretation of flow analyzes that are based on two-particle 
correlations~\cite{Posk98}. In order to investigate further the 
systematic uncertainties due to non-flow, comparison of azimuthal 
correlations is made between Au+Au and p+p collisions, in which only 
non-flow is expected to occur. Additionally, non-flow contributions to 
four-particle correlations, which are expected to be 
small~\cite{STARFlowPRCY1}, are investigated through comparison to 
six-particle correlations.

\vspace{-0.5cm}    

\section{Results}

\begin{figure}[hbt!]
\includegraphics*[height=.25\textheight]{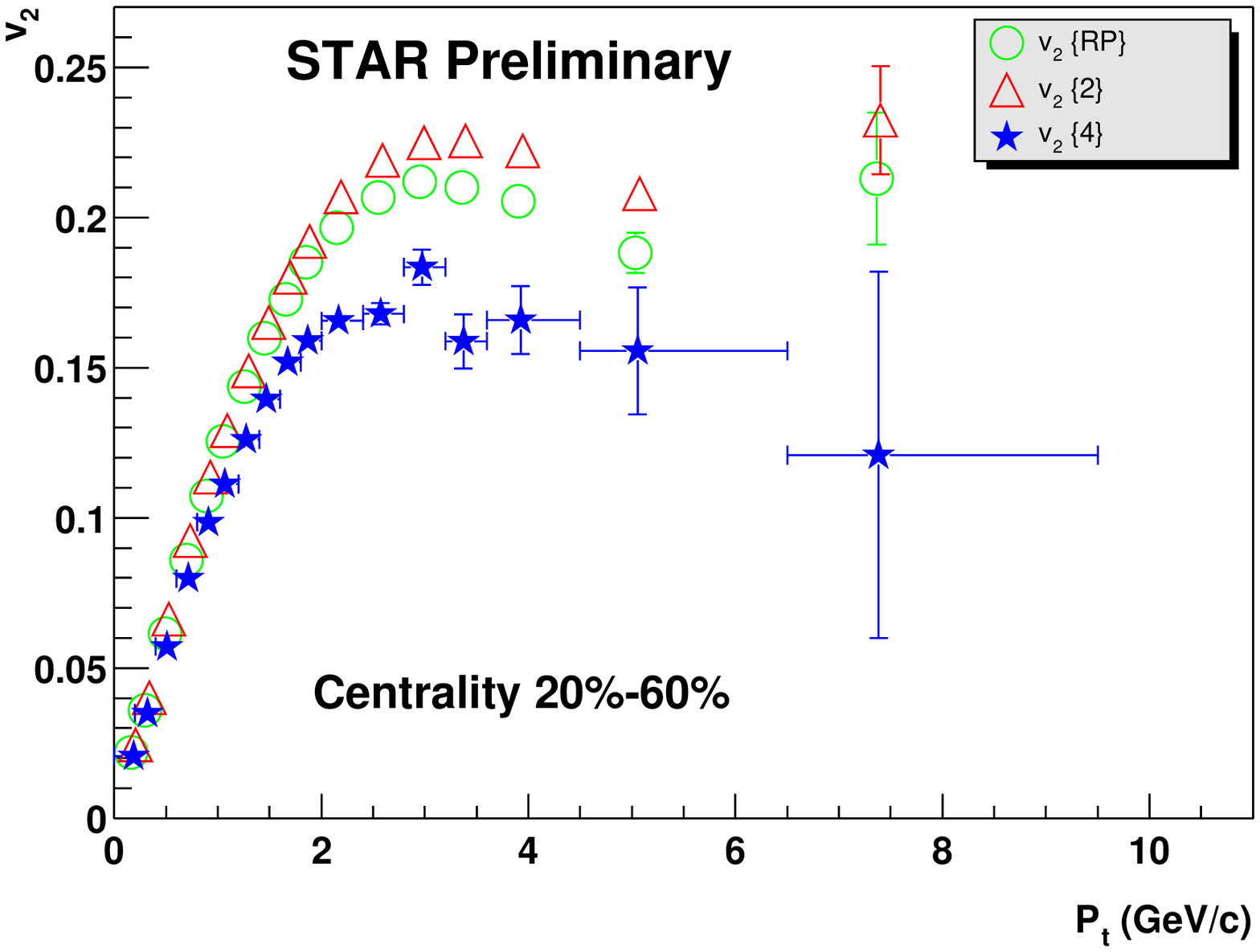}
\includegraphics*[height=.25\textheight]{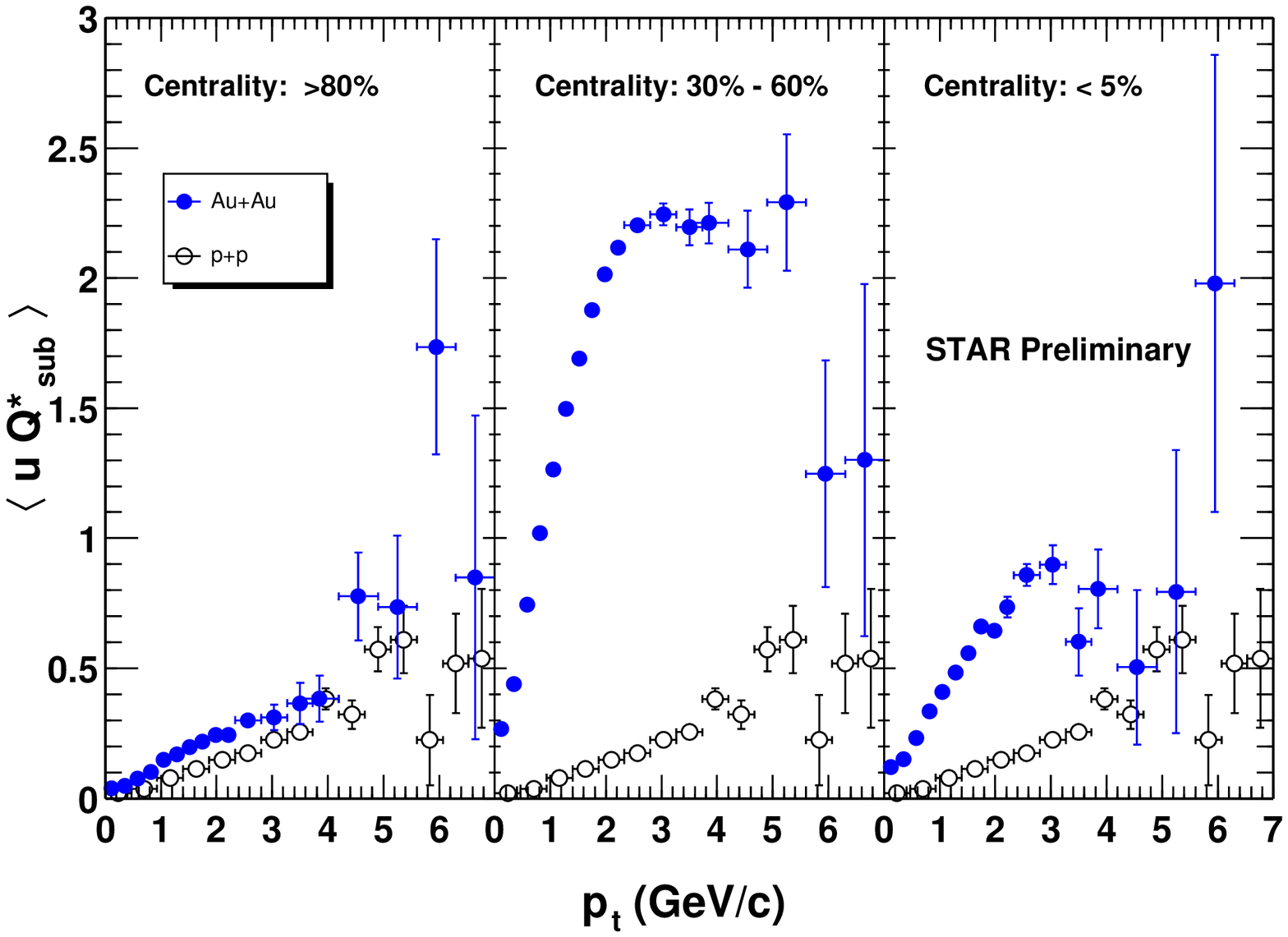}
\caption{\label{fig:vpt_ppAuAu} Left :  $v_2$ as a function of transverse
momentum from event plane method (circles), two-particle cumulant method 
(triangles) and four-particle cumulant method (stars). Right : 
Two-particle correlation in p+p (circles) and Au + Au (solid dots). $Q_{sub}$, which is the $Q$ vector from randomly divided sub-events, is used in making this plot.}
\end{figure}

The left plot of figure~\ref{fig:vpt_ppAuAu} shows $v_2$ from the event plane (RP)
and two- and four-particle cumulant ($v_2\{2\}$ and $v_2\{4\}$) methods. 
The difference between the RP and $v_{2}\{2\}$ methods may originate 
from the analytical extrapolation of event plane resolution 
from sub~-events to full events. The difference between four-particle cumulant $v_2$ and the other 
two methods could be partially explained by non-flow effects 
and partially explained by the fluctuation of $v_2$ 
itself~\cite{MikeRaimondFluct}, but 
in either cases, 4-particle cumulant method gives a lower limit on $v_2$.
From the plot we can see that in middle central events at 
$\sqrt{s_{NN}} = 200$ GeV, significant $v_2$ is found up to 
$7$ GeV/c, which is the region where fragments of high $p_t$ 
partons can be disentangled from the soft hydrodynamics 
component. The measurement will provide an important 
constraint on the initial parton densities in a 
"jet quenching" picture~\cite{MGyulassy2001}. 
The finite $v_2\{4\}$ at high $p_t$ seen in the figure demonstrates 
that $v_2$ is predominantly due to correlations relative to the the 
reaction plane rather than the intra-correlations of 
jet fragments~\cite{Yuri02}."

The four-particle cumulant $v_2$ analysis requires large 
statistics thus has limited power in terms of separation 
of non-flow effects at high $p_t$ with 
currently available data. In order to get an insight to the 
problem, we can separate the two-particle correlation in 
Au+Au collisions as 

\vspace{-0.3cm}    

\begin{equation}       
\langle u_{D} Q^{*} \rangle^{AA} = M ^ {AA} v_D v_I + M ^ {AA} \delta^{AA}_{DI} \, \, ,
\end{equation}       
where $Q = \sum u_j$, $u_j = e ^ {2 i \phi_j}$.  $v_D$ is 
differential flow and $v_I$ is the integrated flow for 
particles used to define $Q$. $\delta^{AA}_{DI}$ is the two-particle non-flow correlation in Au+Au collisions and can be 
approximated to $\frac{\delta^{pp}_{DI}}{N_{collision}} \approx \frac{\delta^{pp}_{DI}M^{pp}}{M^{AA}}=\frac{\langle u_{D} Q^{*} \rangle^{pp}}{M^{AA}} $. $M^{AA}$ and $M^{pp}$ are 
multiplicities for Au+Au collisions and p+p collisions, respectively.

Here we have made assumptions, which are not necessarily true, 
about the similarity of non-flow for both p+p and Au+Au 
collisions, but this comparison is a very useful supplement 
to what is learned from four-particle cumulant 
studies. Rearranging terms, we have

\vspace{-0.4cm}    

\begin{equation}       
\langle u_{D} Q^{*} \rangle^{AA} = M ^ {AA} v_D v_I + \langle u_{D} Q^{*} \rangle^{pp} \, .
\end{equation}

The right plot of figure~\ref{fig:vpt_ppAuAu} shows that the correlations 
have similar magnitude in p+p and the most central and 
peripheral Au+Au collisions, indicating that non-flow may 
dominate the correlations for these centrality classes. 
The correlations for mid-central Au+Au collisions are much stronger 
than those seen in p+p, so that true flow effects may dominate in 
this case.

\begin{figure}[ht]      
\resizebox{20pc}{!}{\includegraphics{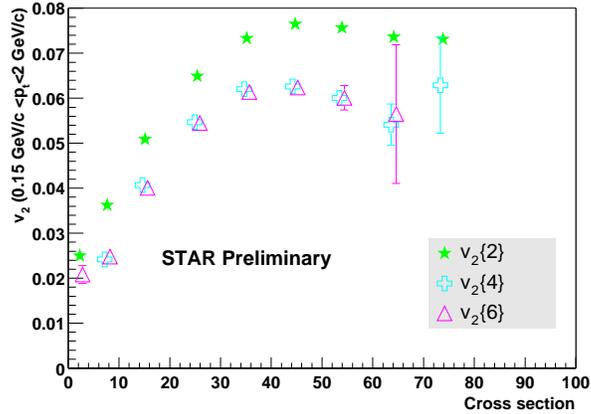}}      
\caption{\label{fig:v246Cent} 
$v_2$ from two-(stars), four-(crosses) and six-(triangles) particle correlation, versus cross section at $\sqrt{s_{NN}} = 200$ GeV  There is an overall $10\%$ downward systematical uncertainty due to low $p_t$ background contaminations in STAR detector. }
\end{figure}

In four-particle cumulant analysis, contributions of non-flow 
effect from four-particle correlations remains although
non-flow from two-particle correlations are removed.   
To remove 4th order non-flow effect one needs to go to
higher order cumulants. However this effect was expected to 
be small~\cite{STARFlowPRCY1} and indeed confirmed by 
Figure~\ref{fig:v246Cent}. In the
Figure $v_2\{4\}$ and $v_2\{6\}$ overlays on the top of 
each other very well, indicating that non-flow from four-particle 
correlation, which is shown by the difference between 
$v_2\{4\}$ and $v_2\{6\}$, is negligible.

\section{Summary }

Sizable $v_2$ is found up to $7$ GeV/c in $p_t$ in Au + Au 
collisions at $\sqrt{s_{NN}} = 200$ GeV. Non-flow effect 
could be dominant at high $p_t$ in peripheral and central 
events. Non-flow from pure four-particle correlation is 
negligible.

\section{Acknowledgment}
I thank S. Voloshin, K. Filimonov, A. Poskanzer and R. Snellings for their contributions.

\end{document}